\documentstyle[12pt]{article}
\begin{document}
\vskip 2. true cm
\begin{center}
{\large \bf On the possibility of decoherence due to relativistic  
 effect} \\
      \vskip 1.4 true cm
        B. Carazza \\
       \vskip .1 true cm
\par 
       {\it Dipartimento di Fisica dell' Universit\`a,} \\
      {\it viale delle Scienze, I43100 Parma, Italy}\\
        {\it INFN Sezione di Cagliari, Italy}\\  
      \end{center}
\par \vskip 2mm
\begin{center} {\bf  Abstract} \end{center}
 \vskip 1mm
\begin{quote}
This note looks at the possibility of a system of free  
particles presenting decoherence in the total momentum
when tracing upon their relative momenta if we
 take into account a relativistic correction
 to the expression of the kinetic energy.
\end{quote}
\vskip 1.5 true cm
Key words: Decoherence, reduced density matrix, preferred basis, 
 isolated system.
\vskip 1.5 true cm
PACS: 03.65.Bz
 \newpage
      \noindent{\bf 1. INTRODUCTION}

\vskip 1.6 true cm
 \par
Decoherence - the phenomenon whereby a quantum system initially in a 
pure state evolves into a improper mixture due to interaction with its 
environment\cite{omn} - has been studied extensively, using both toy 
models\cite{zur} and more complex models representing real physical 
situations\cite{jaz}. Sources of decoherence such as 
scattering\cite{gal,teg} and quantum gravity\cite{hav,ell} and both 
internal\cite{ha,flo}
 and external\cite{cal,joo} environment are taken into 
consideration.
 Omnes\cite{omn2}  recently presented
 a general theory of the effect of
decoherence that includes both the harmonic model of Caldeira-Legett
\cite{cal} and
the external environment considered by Joos and Zeh\cite{joo}
 and is related to the quantum state\cite{qua} diffusion model.
 \par
  Decoherence is especially interesting as it promise to
  solve the old problem of how to derive the classical behaviour
  of macroscopic bodies from quantum principles.
       In fact, in the case of a macroscopic body, decoherence is believed
      to suppress the off-diagonal elements of the spatial 
      reduced matrix of the
      centre of mass, irrespective of initial conditions. 
      This is equivalent to establishing a superselection
      rule for the position of macroscopic bodies and
      to saying that we cannot
      experience spatial macrosuperpositions.
      We can thus explain why we do not encounter states
      of this kind when looking at everyday objects.
      To derive such a result it is not necessary to
      consider an external environment.
      In fact, a system of many particles 
      like a macroscopic body may be considered as consisting
      formally of a  ``collective system''  described by the
      collective coordinates, and the system
      described by its microscopic coordinates,
      which can act as an (internal) environment
      if the two formal systems are coupled.
      This coupling may be either guaranteed by some constraints or
      caused by an external potential.
      Of course in the case of an isolated macroscopic body,
      the Hamiltonian consist of
      the sum of two separate Hamiltonians  relative to the centre
      of mass and internal variables respectively,
      there is no coupling between them
      and so decoherence of the collective variable is not possible.
      However, we wish to signal a possible case of decoherence,
      with suppression of correlations in the total momentum,
      for a system of particles even if isolated,
      provided that we consider relativistic corrections
      to the non-relativistic Hamiltonian.
      It would pointed out that decoherence of isolated
      systems has been discussed before in various
      contexts: Halliwell\cite{ha} investigated the variables that will
      generally become effectively classical as the local densities
      (of number of particles, momentum, energy),
      Calzetta and Hu\cite{calz} the decoherence of histories of certain
      correlation functions in a field theory.
      We should also mention that the diagonalization in the
      square of momentum  basis has been argued for a
      system of (non-relativistic) particles interacting 
      with the linearized gravitational field\cite{ana}.
 \par
      It is know that in classical electrodynamic a system
      of interacting particles may be approximately described
      to terms of second order in $1/c$ by a Hamiltonian  
      depending only on their positions and velocities,
      since the radiation comes in  with the terms of order $1/c^3$.
      The same may be said for particles subject to gravitational
      interaction\cite{lan}. In this approximation the
      total momentum of the system
      is no longer independent, but is coupled to
      the internal degrees of freedom.
      So, if we adopt the above mentioned Hamiltonian (after
      due symmetrization) 
      as the Hamiltonian operator within non-relativistic
      quantum mechanics, we may expect to have decoherence
      for the collective variable tracing over the
      microscopic ones. 
      This may occurs even if the particles are not
      interacting, thanks to the relativistic correction
      to their kinetic energy.
      The coupling we are talking about commute
      with the non-relativistic free Hamiltonian
      of the centre of mass (c.m.).
      As we learned from studies of wide-open
      quantum systems\cite{per}, the diagonalization is to
      be expected in the basis of any dynamical operator commuting
      with the coupling terms and so your candidate
      to be a preferred basis is that of total momentum.
      The importance of an interaction commuting with
      the preferred basis was already stressed
       by Zurek\cite{zurr}.
 \par
      Since the aim of this note is merely to briefly
      signal such a possibility,
      we will consider for the sake of simplicity a 
      system of $n$ free spinless particles of equal mass $m$ in one
      dimension, a one dimensional ideal gas in fact.
      We stress that we intend to work within the non-relativistic
      quantum mechanics and consider the correction to the kinetic
      energy and its consequences as a relativistic effect.
 \par
      The model certainly is not realistic but we intend to
      offer the result as a further case of decoherence
      with not usual features. Not usual because the system is
      isolated and also because in our case the reduced statistical
      operator diagonalizes on the basis of the total momentum,
      which thus acquires the status of preferred basis.
      We believe all this is interesting in itself.
      We do like to stress that the calculations will be
      carried out without recourse to any master equation but
      by following the Schroedinger evolution of a pure state.
      With the kinetic operator used, the Galilean invariance
      is destroyed, so we have to specify the reference frame
      we are referring to. A Galilean transformation corresponds
      to a translation of momenta. For the approximation we
      are using to make any sense, the reference frames to
      take into consideration are obviously the ones (if they
      exist) for which the absolute value of momentum for
      each particle is not significantly likely to be of
      the order of or greater than $mc$.
      This requisite can be expressed by applying the
      condition $\langle \widehat p_i^2 \rangle \ \ll m^2c^2$ for every $i$.
      It will certainly be satisfied if the more restrictive
      condition $\sum_{i=1}^n  \langle \widehat p_i^2 \rangle  \ \ll m^2c^2 $
      also is. To establish a definite reference frame, we
      will adopt the one in which, for a given wave function,
      the last mentioned expression gives the lowest value.
      It will be the reference frame in which the mean
      value of total momentum is zero. Even selecting the optimal 
      reference frame, however, the validity of our approximation
      is not automatically guaranteed and our result will
      be only significant if the condition on the values
      of $ \langle \widehat p_i^2 \rangle $ is satisfied.

 \vskip .9 true cm

 \noindent{\bf 2. THE SYSTEM OF FREE PARTICLES IN ONE DIMENSION
   WITH RELATIVISTIC CORRECTION TO THE KINETIC ENERGY }
 \vskip .6 true cm
 \par
    The free Hamiltonian which we will consider 
    for quasi-relativistic particles is:
 \begin {displaymath}
 \sum_{i=1}^n { \widehat p_i^2 \over 2m} - \sum_{i=1}^{n} { \widehat p_i^4 
    \over 8 m^3 c^2}
  \qquad .
 \end{displaymath}
  Using the momentum basis and with $\varphi(p_1, p_2,\ldots,p_n)$
 indicating the initial state, we have the state at time $t$
 as follows:
 \begin {displaymath}
  |\chi(t)\rangle  = e^{-it/\hbar [ \sum_{i=1}^n ({ p_i^2 \over 2 m} -
  { p_i^4 \over 8 m^3 c^2} )]} \, \varphi(p_1, p_2,\ldots,p_n) \qquad .
 \end{displaymath}
      We now change the variables: $ p_i = P/n+\eta_i$,
      where $P$ is the total momentum.
      The $\eta_i$  are not independent, since they must satisfy
      the relation $\sum_{i=1}^n \eta_i = 0$.
      We will consider as independent variables $P$ together with 
      the first $n-1$  relative momenta.      
      The last relative momentum
      is expressed as $\eta_n = -\sum_{i=1}^{n-1} \eta_i$.
      The Jacobian determinant of the
      transformation is equal to $1$.
      We will write the integration volume element for the new
      variables as $dV = dP \, dS $, where
      $dS = d\eta_1 \, d\eta_2 \, \cdots \,  d\eta_{n-1}$.
      Using the new variables and taking into account the
      relation $ \sum_{i=1}^n \eta_i = 0$
      the expression above now becomes:
 \begin  {displaymath}
  |\chi(t)\rangle  =
  e^{-{it \over \hbar} [{P^2 \over 2nm} - {P^4 \over 8n^3m^3c^2}
 + \sum_{i=1}^n( {\eta_i^2 \over 2m}     -  {\eta_i^4 \over 8m^3c^2} 
  -{3P^2 \over 4n^2m^3c^2} \eta_i^2  -{P \over 2nm^3c^2}  \eta_i^3 )] } \,
\Phi(P, \eta_1,\ldots,\eta_n)   
 \end  {displaymath}
 where $\Phi(P, \eta_1,\eta_2,\ldots,\eta_n)$ indicates 
 $\varphi(P/n +\eta_1, P/n+\eta_2,\ldots,P/n+\eta_n)$
 Note that here and 
 in the various expressions that follow 
 $\eta_n$ stands for  $ = -\sum_{i=1}^{n-1} \eta_i$.
 The projection $\langle P'| \chi(t)\rangle$ of the state 
vector at time $t$ on an eigenstate $|P'\rangle$ of the total momentum is:
\begin {displaymath}
 e^{-{it \over \hbar} [{P'^2 \over 2nm}-{P'^4 \over 8n^3m^3c^2}
 +\sum_{i=1}^n( {\eta_i^2 \over 2m}-{\eta_i^4 \over 8m^3c^2} 
  -{3P'^2 \over 4n^2m^3c^2} \eta_i^2 -{P' \over 2nm^3c^2} \eta_i^3)]} \,
 \Phi(P',\eta_1,\ldots,\eta_n) .   
 \end {displaymath}
By tracing $\langle \chi(t)|P''\rangle \langle P'|\chi(t)\rangle$ upon the 
relative momenta 
 i$.$e.\ integrating over $dS$ we get the reduced density matrix
 element on the total momentum 
basis:
 \begin {displaymath}
 \rho_{_{P'P''}}= e^{{-it \over \hbar} [{ (P'^2-P''^2) \over 2nm} - 
 {(P'^4-P''^4) \over 64 m^3 c^2}]} \, I_{P'P''} \qquad .
 \end{displaymath}
  The factor $I_{P'P''}$ indicates the integral
 \begin {displaymath}
 I_{P'P''} = \int \Phi(P', \eta_1,\eta_2,\ldots,\eta_n) \,
 \Phi^{\ast}(P'', \eta_1,\eta_2,\ldots,\eta_n) \,
  e^{it  \sum_{i=1}^n ( w \eta_i^2 + z \eta_i^3 )} \, dS 
\end{displaymath}
 where $w = 3 (P'^2-P''^2)/(4 \hbar n^2 m^3 c^2) $ and
 $z =  (P'- P'')/(2 \hbar n m^3 c^2) $.  
 The diagonal matrix elements are constant of motion,
 but when $P'\neq  P''$ the integral in the previous
 expression  may cancel out with time, in which case the
 off-diagonal elements are cancelled out, meaning
 that we have decoherence.
 \vskip .9 true cm

 \noindent{\bf 3. EVALUATION OF THE REDUCED DENSITY MATRIX ELEMENTS}
 \vskip .6 true cm
 \par
      In order to understand the time behaviour of the 
      matrix elements we are interested in is thus necessary
      to evaluate the integral $I_{P'P''}$. To take into
      account the constraint $\eta_n = -\sum_{i=1}^{n-1} \eta_i$
      we multiply the integrand by the delta function 
      $\delta (\sum_{i=1}^{n} \eta_i) $ using its Fourier 
      representation $ (1/2\pi)\int e^{i \sum_{i=1}^{n} k \eta_i} dk $
      and consider $\eta_n$ too as a independent integration
      variable.      
      Writing 
 \begin {displaymath}
  G(P', P''; \eta_1, \ldots,\eta_n)= \Phi(P', \eta_1,\eta_2,\ldots,\eta_n) \,
 \Phi^{\ast}(P'', \eta_1,\eta_2,\ldots,\eta_n)
 \end{displaymath}
 and $ h(\eta_i) = w \eta_i^2 + z \eta_i^3 $ 
 our integral becomes      
 \begin {eqnarray}
 I_{P'P''} =& (1/2\pi)\int dk \int e^{ik\eta_1+it h(\eta_1)}d\eta_1
 \int e^{ik\eta_2+it h(\eta_2)}d\eta_2  \ \ldots \nonumber \\  
& \int e^{ik\eta_n+it h(\eta_n)}d\eta_n  \, G(P', P''; \eta_1,\eta_2, 
 \ldots,\eta_n)  \qquad . \nonumber
 \end{eqnarray}
  Let us start with the integration on $\eta_n$. We are interested in 
      the behaviour of the off diagonal matrix elements at
      large time, so we shall use asymptotic
      expansions methods to evaluate the result. More
      precisely, we will look at the method of stationary phase\cite{erd}.
      According to which,  when $t$ is a large 
      quantity, the major contribution to the value of the integral
      arises from the immediate vicinity of the end points
      and from the vicinity of those point at which $h(\eta_n)$
      is stationary, i$.$e.\ its first derivative $h'(\eta_n) $
      is equal to zero.
      In the first approximation the contribution of
      stationary points is more important.
      If we suppose that the function representing the intial state
      is square integrable and thus going to zero
      at infinity, we can disregard the end points
      contribution at all. 
      Since only the neighbourhood matters, asymptotic evaluation consists
      of replacing $h(\eta_n)$ near any stationary point     
      $\eta_n^{\ast}$ as  $h(\eta_n^{\ast})
      + (1/2)h''(\eta_n^{\ast})(\eta_n - \eta_n^{\ast})^2$.
      By the same argument, the factor of $ e^{it h(\eta_n)}$
      in the integrand, assuming it is
      a continuous function, is replaced by its value at $\eta_n^{\ast}$
      and the integration is safely extended from $-\infty$ to $+\infty$.
      We will replace $G(P', P''; \eta_1, \ldots,\eta_n)$
      by $G(P', P''; \eta_1, \ldots,\eta_n^{\ast})$, but 
      maintain the dependence on $\eta_n$ of $ e^{ik\eta_n}$,
      since it belongs to the Fourier representation of
      the delta function, a non-continuous and singular 
      function indeed.
      We have two stationary points 
      at $\eta_n = 0 $ and $ \eta_n = -2w/(3 z) = -(P'+P'')/2 $
      These two values are indicated as
      $\alpha_1$ and $\alpha_2$ respectively.
      The corresponding values of the second derivative
      are $ h''(\alpha_1) = 2w $ and $h''(\alpha_2) = -2w $.
      Using the procedure illustrated above,
      we obtains:
 \begin {eqnarray}
   &G(P', P''; \eta_1,\eta_2, \ldots,\alpha_1)[2i\pi/th''(\alpha_1)]^{1/2}
   e^{ith(\alpha_1)+ik\alpha_1-ik^2/(2th''(\alpha_1))}  \ + \nonumber \\
   &G(P', P''; \eta_1,\eta_2, \ldots,\alpha_2)[2i\pi/th''(\alpha_2)]^{1/2}
   e^{ith(\alpha_2)+ik\alpha_2-ik^2/(2th''(\alpha_2))}  \qquad . \nonumber
 \end{eqnarray}
      Integrating now on $\eta_{n-1}$ the two terms of
      the expression above we obtain the sum of four terms,
      which are again doubled after the next
      integration and so on. Lastly the asymptotic
      expression of $I_{P'P''}$ become 
 \begin  {displaymath}
 {1 \over 2 \pi} \int dk \sum_{i_1 i_2 \ldots i_n} 
 G(P', P'';\alpha_{i_1},\alpha_{i_2}, \ldots, \alpha_{i_n})
 \prod_{j=1}^n  \sqrt{{2i\pi \over th''(\alpha_{i_j})}}
  e^{ it h(\alpha_{i_j})+i[k \alpha_{i_j}-{k^2 \over
  2t h''(\alpha_{i_j})}]} 
 \end  {displaymath}
      In the multiple sum, each of $ i_1, i_2, \ldots,i_n $ assumes
      the value 1 or 2, corresponding to the two values
      of the stationary points for each relative momentum.
      Finally, performing the integration
      on $k$, we obtain:
 \begin  {displaymath}
  ({2\pi \over t})^{{(n-1) \over 2}} \sum_{i_1 i_2 \ldots i_n}\hspace{-3mm}' 
 G(P', P'';\alpha_{i_1},\alpha_{i_2},\ldots,\alpha_{i_n})
{e^{{it (\sum_{j=1}^n \alpha_{i_j})^2 \over 2 \sum_{j=1}^n 
1/h''(\alpha_{i_j})}} \over  \sqrt{ i \sum_{j=1}^n 1/h''(\alpha_{i_j})}}
 \prod_{j=1}^n {e^{ it h(\alpha_{i_j})} \over \sqrt{-ih''(\alpha_{i_j})}}
 \end  {displaymath}
 where $\sum'$ means that the terms
 for which $\sum_{j=1}^n 1/h''(\alpha_{i_j})=0$ are excluded.
      Such terms indeed give zero after the integration on $k$.
      After substitution of  
     $ |\sqrt{ i \sum_{j=1}^n 1/h''(\alpha_{i_j})}|$
      with the lower possible value $\sqrt{|1/(2w)|}$,  we have asymptotically
 \begin  {displaymath}
  | \rho_{_{P'P''}}| \leq |{2\pi\over t}|^{(n-1)/2}|2w|^{-(n-1)/2}
  \sum_{i_1 i_2 \ldots i_n}\hspace{-3mm}' \
  |G(P', P'';\alpha_{i_1},\alpha_{i_2}, \ldots, \alpha_{i_n})| \quad .
 \end  {displaymath}
 For the primed sum we may also write:
 \begin  {eqnarray}
&\sum_{i_1 i_2 \ldots i_n}^{\, \prime}\ 
  |G(P', P'';\alpha_{i_1}, \ldots, \alpha_{i_n})| \leq   \qquad \qquad \qquad  \qquad \nonumber \\
  & 2^{n}[ \sum_{i_1 i_2 \ldots i_n}^{\, \prime} \
 |\Phi(P', \alpha_{i_1},\ldots,\alpha_{i_n})|
 | \Phi^{\ast}(P'', \alpha_{i_1},\ldots, \alpha_{i_n})|]/p     \nonumber
 \end  {eqnarray}
 where $p$ is the number of terms which contribute with
 non zero value to the sum.
  We used the definition of
 $ G(P', P''; \eta_1,\eta_2, \ldots,\eta_n)$
  already given and the fact that $p$ is equal or less than
 the $2^{n}$ possible combinations of the indexes $i_j$.
 \par
      At this point we will compare the off-diagonal
      matrix elements with the diagonal ones.
      Since the last are constant of motion, we
      do not expect it would change the result, but
      to obtain a dimensionless expression.
      To this end we need to establish a further bound.
      Looking at the definition of
      the reduced matrix elements we may state that
  \begin{displaymath}
    | \rho_{_{PP}}|=I_{PP} = \int \Phi(P, \eta_1,\ldots,\eta_n)
 \Phi^{\ast}(P, \eta_1,\ldots,\eta_n) \,  dS
    \end{displaymath}
      is evidently greater or equal to the result of
      the integration over a finite volume as the
      integrand is a positive quantity.
      Let $ -Mc \le \eta_i \le Mc $ be a finite integration
      interval for each variable. Recalling that
      only $n-1$ of them are independent, as
      $\eta_n = -\sum_{j=1}^{n-1} \eta_j$:
   \begin  {displaymath}
  |\rho_{_{PP}}| \ge (2Mc)^{n-1} \,  |\Phi(P, u_1, u_2 \ldots )|^2
 \end  {displaymath}
    where $ u_1, u_2 \ldots $ denotes an internal point  
      of the hypercube  chosen as integration domain.
 We recall that the function $|\Phi(P, u_1, u_2 \ldots )|$
 is assumed as bounded everywhere.
     Comparing now  the off-diagonal elements
       to the diagonal ones
      we will consider, to be democratic, the ratio of
  $| \rho_{_{P'P''}}|$ to $ (|\rho_{_{P'P'}}| | \rho_{_{P''P''}}|)^{1/2}$.
 Using the inequalities already found:
 \begin  {displaymath}
 | \rho_{_{P'P''}}|/| (\rho_{_{P'P'}}|| \rho_{_{P''P''}}|)^{1/2}
   \le  2 A |t/\tau|^{-(n-1)/2}  \qquad .
 \end  {displaymath}
 The constant $A$ denotes the quantity:
 \begin  {displaymath}
  {[ \sum_{i_1 i_2 \ldots i_n}' \
 |\Phi(P', \alpha_{i_1},\ldots,\alpha_{i_n})|
 | \Phi(P'', \alpha_{i_1},\ldots, \alpha_{i_n})|]/p \over
 |\Phi(P', u'_1, u'_2 \ldots )||\Phi(P'', u''_1, u''_2 \ldots )|}
 \end  {displaymath}
 and $\tau= {4 \pi \hbar m \over 3 |P'^2-P''^2|}$
   is a characteristic time scale
      of the process such that when $t \gg \tau$ the absolute values
      of the off diagonal elements are vanishing.
   It would be pointed out that
   actually we have decoherence in the basis of
  (non-relativistic) kinetic energy $ E = {P^2 \over 2M} $
   of the c.m.. Introducing $ E $ as 
   a more appropriate
  variable the decoherence time is better written
  as  $\tau=  {2 \pi \hbar m \over 3 M |E'-E''|}$.
  It is inversely proportional to the difference
  $| E'-E''|$
  and  also decreases with increasing 
  total mass $M$ as one may expect intuitively.
      With $| E'-E''|$  of the order of a few 
      $eV$ and with, say, ten particles,
      the characteristic time is very small,
      about $10^{-16} sec$.
\par     
 As we anticipated and contrary to our initial prediction,
 the result we have 
just found shows that  the reduced density
 matrix diagonalizes on the basis of $\widehat P^2$ and not
 on that of total momentum $\widehat P$.
 That is, they are the matrix elements for which
 $|P'| \neq |P''|$, which we found go to zero as time increases.
 But our expression for the asymptotic behaviour is not valid 
if $ P'= -P''$. In such a case, in fact, the function $h(\eta)$
 depends only on $\eta^3$, we have only a stationary point of
 the second order and the asymptotic evaluation of our 
integral has to be worked out separately. To this end, the integration
 interval in $\eta_i$  is split into $ (0, \infty)$ and $(-\infty, 0)$
 and each of the two integrals is integrated by parts a number of times,
 differentiating $G(P', P''; \eta_1,\eta_2, \ldots,\eta_n)$ and
 integrating the remaining factor of the integrand. Taking only the
 dominant term into account, asymptotically we have:
 \begin{displaymath}
   | \rho_{_{P'P''}}|/  (|\rho_{_{P'P'}}| | \rho_{_{P''P''}}|)^{1/2}
  \leq B \,  |t/\tau'|^{-(n-1)/3}| \int^{+\infty}_{-\infty} d\beta 
 [A_i(\beta)]^n |  \qquad .
 \end{displaymath} 
 The constant $B$ is:
 \begin{displaymath} 
 {|\Phi(P', 0,0,0,\ldots,0)| \, | \Phi(P'',0,0,0,0,\ldots, 0)|
 \over |\Phi(P', u'_1, u'_2 \ldots )| \, 
 |\Phi(P'', u''_1, u''_2 \ldots )|}
 \end{displaymath} 
  and $\tau'={ 2 \pi^3  \hbar \over 3 n^2 |P'-P''| c}$.
 For values of $|P'-P''| $  of a few $eV/c$ and always 
 assuming  $n \sim 10$, still $\tau' \sim  10^{-16} sec$.
 The function $ Ai(\beta)$ indicates the Airy function
 $ {1 \over \pi} \int^{+\infty}_0 cos(\beta y +y^3/3) dy $.
 We assumed time to be positive
 and $P' > P''$. The asymptotic estimate of the other half
 of the off-diagonal matrix elements for which $P' < P'' $
 is immediately obtained since
 $|\rho_{_{P''P'}}| = |\rho_{_{P'P''}}| $ 
 by hermiticity.
 $ Ai(\beta)$ for large and positive values of $\beta$ 
 tends exponentially to zero. It is also going to zero for
 $\beta$ which tends towards $ -\infty $,
 but now the behaviour is like $|\beta|^{-1/4}$. However, for a 
sufficient number of particles (five or more) $[Ai(\beta)]^n$
 is integrable. Summing up,
 $| \rho_{_{P'P''}}|/ (|\rho_{_{P'P'}}| | \rho_{_{P''P''}}|)^{1/2}$,
 if the number of particle
 is as indicated, also tends asymptotically to zero when $P'=-P''$
 and we have a complete diagonalization on $\widehat P$ basis.
 \par
 These findings are strongly dependent on the assumption that
 the system of particles is isolated. It presumes a not very realistic
 situation since from physical viewpoint it is hard to
 exclude external influences,
 and even under the slight perturbation the time behaviour
 we obtained for the reduced matrix elements would 
 not persist. So, even if we intended to present
 our findings strictly as a mathematical result, a brief
 discussion about the persistence of the present decoherence
 effect in the case of external perturbations imposes.
 We will discuss this point in comparison with the case
 of a free non-relativistic particle.
 \par
 The time dependence in the momentum basis of a
  non-relativistic free particle density matrix 
 elements is simply:
   \begin{displaymath}
\psi(k') \,  \psi^{\ast}(k'') \, e^{-it \, {{(k'^2} - 
 k''^2) \over 2 m \hbar}}
      \end{displaymath}
 where $\psi(k)$ is the initial state.
 Under a time average the expression above, compared to
 the diagonal elements, which are constants of motion,
 becomes:
  \begin{displaymath}
 {\varrho_{_{k' k''}} \over  \varrho_{_{k' k'}}^{1/2} \,  \varrho_{_{k'' k''}}^{1/2}} =
 {\psi(k')  \, \psi^{\ast}(k'') \over |\psi(k')| |\psi(k'')| } \,  {1 \over t} 
 \int^t_0 e^{-it/\tau_{_{free}}} dt
      \end{displaymath}
 with, denoting as $T_{k}$ the kinetic energy,
 a ``decoherence time'' $\tau_{_{free}} = { \hbar \over 
 |T_{k'}-T_{k''}| } $.
 Since the integral gives a limited oscillating term
 the asymptotic behaviour for the correlations of two different
  eigenstates of kinetic energie is as the inverse power of time, namely
 as $ (t/\tau_{_{free}})^{-1}$.
 Let now consider a constant perturbation given as 
 a generic function of the energy $ \varepsilon f(T) $, which commutes
 with the free Hamiltonian, and defines $ T_{k''}=T_{k'}- \Delta$.
 \par
 Taking the time average requires to substitute 
 the former integral with the new expression:
\begin{displaymath}
 {1 \over t} \int^t_0
 dt \  e^{- {it \over \hbar} \,  [ \Delta + \varepsilon f(T_{k'}) - 
  \varepsilon f(T_{k'} - \Delta)] } .
 \end{displaymath}
 For each value of $ \Delta $ for which
 the argument of the exponential is zero the integral gives
 just one. Then for each $k'$ there are so many values
 of $k''$ as the zeroes of the function $ f(T_{k'}) -f(T_{k'} - \Delta) $
 for which
 the averaged matrix elements do not go to zero with
 increasing time.
 In the case of decoherence we considered the role of a free particle is
 played by the centre of mass. Adding any constant perturbation
  $\varepsilon f(T_{P})$ to the Hamiltonian of our isolated system 
  will not disturb at all the asymptotic behaviour of
  any reduced matrix element. In fact we merely have a
 change of their phases, whereas their absolute values
 are left unchanged. This is a rather artificial consideration,
 since the sought disturbance is not very realistic.
  However this example may deserve
  to show that our case of decoherence is persistent under
  disturbances which only
 touches the matrix elements phases.
 \par
 Coming back to the case of the free non-relativistic
 particle, we now consider a perturbation 
 due to a constant force, adding to the free Hamiltonian
 the term $H'= x F $, which
 may be written as $ x m c/\tau_l$. Choosing 
 the acting force as the weight force,
 the quantity $ \tau_l $ is about $ 3 \times 10^7 sec$.
 \par
 It is easy to see that the wave function of the free falling particle
 evolves with time in momentum basis as:
  \begin{displaymath}
 \psi(k + m  c t/\tau_l) \, e^{-{i \over 2 m \hbar } ( k^2 t +  k {m c \over \tau_l} t^2 + {m^2 c^2
 \over 3 \tau^2_l} t^3)}
    \end{displaymath}
 with the same initial condition as before.
 The above expression means that in addition to the phase
 variation with time of each plane wave component,
 the wave packet ``moves'' in momentum
 space at the rate of $mc/\tau_{l}$ per $sec$.
  This entails that, for a square integrable wave function,
  $\psi(k+mct/\tau_{l})$ goes in the long run to zero for
  any fixed value of $k$ and hence any density
  matrix element with fixed indexes do the same.
  For a wave packet with initial spread $ \Delta k $
  equal to some fractions of $mc$ it will happen after
  a time of the order of year.
  So we may disregard this kind of time dependence for
  the first minutes. Within this approximation the time average
  of the normalized density matrix elements gives:
        \begin{displaymath}
{\psi(k')\,\psi^{\ast}(k'') \over |\psi(k')| \, |\psi(k'')|} \, {1 \over t} \int^t_0 
 e^{-{it \over 2 m \hbar} (k'^2-k''^2)-i{m c (k'-k'') \over 2 m \hbar \tau_l} t^2} \, dt
    \end{displaymath}
 We rewrite the time factor as
  \begin{displaymath}
{ 1 \over t} \int^t_0 e^{-i({t \over \tau_{_{free}}} +{t^2 \over \tau^2_m})} \, dt
   \end{displaymath}
 where $\tau_{_{free}}={ \hbar \over |T_{k'}-T_{k''}|}$ was
 already encountered and 
$ \tau_{m} ={|k'+k''|  \over  m c} (\tau_{_{free}}  \tau_{l})^{1/2}$.
  We point out that $\tau_{m}$ is generally greater
 than $\tau_{_{free}}$.
 The asymptotic expansion gives:
 \begin{displaymath}
{1 \over t}  \int^t_0 e^{-i({t \over \tau_{_{free}}} +{t^2 \over \tau^2_m})} \, dt \,
 \simeq \,{ e^{-it/ \tau_{_{free}}} \over t}  \, \int^t_0 e^{-i { t^2 \over \tau^2_m}} \, dt 
 \, + \, \cdots
    \end{displaymath}
 The first dominating term is easy written  with a couple of
 Fresnel integrals $ C(z)$ and $S(z)$ where $z= \sqrt{2/\pi} \, t/\tau_m$:
   \begin{displaymath}
 { e^{-it / \tau_{_{free}}} \over t} \,  \int^t_0 e^{-i { t^2 \over \tau^2_m}} \, dt \,
=  \, {\sqrt{2/\pi} \, e^{-it/\tau_{_{free}}} \over t/\tau_m} \, [ C  (\sqrt{2/\pi} \, t/\tau_m) -
i S (\sqrt{2/\pi} \, t/\tau_m)]
     \end{displaymath}
 For  $ z < 1$ the Fresnel integrals 
 are increasing functions of the argument, and both of them
 goes to the constant value  $1/2$ in the limit of large $z$.
 In conclusion in the case of non-relativistic particle
 subjected to weight force we have still decoherence
 with a $1/t$ asymptotic behaviour.
 The decoherence time however is now the greater quantity $\tau_{m}$
 instead of $\tau_{_{free}}$.
 We also examined our system of free quasi-relativistic
 particles, in the case $|P'| \neq |P''|$,
  under the perturbation of the weight force
 applied to the c.m.. We encountered again three characteristic
 time scales, one of which is the already defined $\tau$.
 At time  $t \gg \tau$ but shorter than the analogue
 of $\tau_{m}$ and in the worse situation the off-diagonal
 density matrix elements differ from zero.
 We have still decoherence, but with a greater 
 decoherence time, the analogous of  $\tau_{m}$.
 However the time behaviour
 for the decays of correlations is now as $1/t^{n-1}$.
 \par
 We may conclude that the decoherence mechanism
 we presented here is as robust or a little more
 robust to perturbations than the non-relativistic
 free particle case, i. e. not very robust.
 \par
 For the sake of completeness,
 we will also consider the case of relativistic particles,
  using the so-called  ``relativistic Schroedinger equation''.
 This equation has been used in various
 situations, for
 example to study the stability of matter constituted of fixed nuclei
 and electrons and 
 the collapse of stars of self-gravitating bosons or fermions\cite{car}.
 It is obtained replacing in the usual Schroedinger equation
 the kinetic energy of a particle with its full relativistic expression.
 Let us suppose that there
 is a Lorentz frame such that $ \sqrt{ \langle \eta_i^2 \rangle_{_P}} 
  \ll |P/n| $ for a set $R$ of values of
 $P$, the average value in question being calculated by means of the 
 above mentioned function
 $\Phi(P, \eta_1,\eta_2,\ldots,\eta_n)$. In these conditions we can use
 a procedure similar to
 the previous one to calculate the time behaviour of the off-diagonal
 reduced density matrix
 elements whose indexes $P', P''$ refer to eigenvalues of the total impulse 
 belonging to the 
 set $R$. In fact, the kinetic energy $\sqrt{m^2c^4 + (P/n +\eta_i)^2}$ 
 of each particle $i$ can be developed with good approximation
 in power of $\eta_i$ retaining the
 terms up to $\eta_i^3$. Repeating, mutatis mutanda, the previous steps 
 we once again find
 that the off-diagonal elements of the reduced density matrix
 compared to diagonal ones decrease 
 asymptotically with
 time, as follows:
 \begin  {displaymath}
 | \rho_{_{P'P''}}| / (|\rho_{_{P'P'}}| | \rho_{_{P''P''}}|)^{1/2}
   \le  2 C \,  |t/\tau_r|^{-(n-1)/2}  \qquad .
 \end  {displaymath}
 The constant $C$ is:
 \begin  {displaymath}
 {[\sum_{i_1 i_2 \ldots i_n}' \
 |\Phi(P', \gamma_{i_1},\gamma_{i_2},\ldots)| \, 
 | \Phi^{\ast}(P'', \gamma_{i_1},\gamma_{i_2},\ldots)|]/p \over
 (|\Phi(P', v'_1, v'_2 \ldots )| \, |\Phi(P'', v''_1, v''_2 \ldots )|)^{1/2}}
 \end  {displaymath}
 denoting as $\gamma_1, \gamma_2$ the two new stationary points.
 The decoherence time is now
 \begin {displaymath}
  \tau_r = {\pi \hbar E'^3 E''^3 \over n M^4 c^8 |E''-E''|(E''^2+
 E'E''+E''^2)}   
 \end {displaymath}
where $E^2=M^2 c^4+ P^2 c^2 $.
For $ E' \sim E'' \sim 10Mc^2$, $|E'-E''| \sim 1eV$ and considering 
about ten particles, we obtain $\tau_r \sim 10^{-13} sec $.
 All of this applies, we remind, if both $|P'|$ and $|P''|$ 
belong to $R$. As before, the result is that the reduced matrix 
diagonalizes on the basis 
of $\widehat P^2$. But here too our expression for the asymptotic behaviour 
is not valid 
if $| P'|= | P''|$. In that case we have a double stationary point, 
the evaluation of the off diagonal reduced matrix elements has
 to be worked out separately with the result that
 asymptotically $ | \rho_{_{P'P''}}|$ 
depends on time as $|t|^{-(n-1)/3}$.

 \vskip .9 true cm
 \noindent{\bf 4. CONCLUDING REMARKS }
 \vskip .6 true cm
 \par
 Using non-relativistic quantum mechanics, and within a reference frame
 in which the system
 is on average at rest (i.e. its average total momentum is zero), we have
 shown that a
 one-dimensional system of a sufficient number of free
 quasi-relativistic particles, even if isolated, may
decohere on the basis of the total momentum if we
considers correction to kinetic energy to terms of second order in $1/c$.
 In other words, we have established, under the conditions stated above,
 a superselection rule for the total
 momentum.
 For the sake of completeness, we obtained 
similar results using the
 full relativistic expression for the kinetic operator for those Lorentz
 frames in which the
 relative momenta are much smaller than $P/n$, where $P$ is the total momentum. 
The decoherence we are talking about
 was obtained tracing upon the relative
 momenta.
 Our choice of that coarse-graining seems natural 
if we want to look at the system
 as a whole.
The higher the number of particles, the faster the decoherence, 
meaning that 
it becomes more effective in the macroscopic limit.
 What we found in the case of relativistic corrections within a 
 non-relativistic quantum mechanics framework may reasonably
 suggest that isolated systems of charged or gravitational
 interacting particles, for which the approximate Hamiltonian we mentioned 
 is adopted, present decoherence on
 the basis of total momentum if we look at the system as a whole and 
ignore the internal degrees of freedom. Systems of this sort might include 
atoms with many
 electrons, for instance. True, the Hamiltonian in question is not only 
approximated but 
also ignores spin, so it may be considered ``semi-classic''. 
The presumed effect is interesting anyway.
 If the number of particles is very big, such that the 
system may be considered
 macroscopic, it cannot be supposed to be isolated, due to the 
extraordinary density of its
 energy levels and to the action of surrounding objects, even if this 
tends towards zero. Also we learned that our decoherence mechanism
 is not bery robust to perturbations and so other non relativistic
interaction terms will cause a more effective and different
phenomenologically decoherence of the c.m. variables.
 The action of the external environment will certainly prevail, 
so that the reduced
 density matrix for the collective variable of our 
system will diagonalize on the basis 
of the position, 
as shown in many examples\cite{joo}.
 But a mesoscopic object, with a very small 
number of components,
 may probably be considered isolated, if only for a short period of time. 
In which case we get 
the following situation: macroscopic bodies appear in a localized state, 
while particles made
 up of a small number of components would appear in well defined states 
of the total momentum.
 Further, while not being realistic, the model we have discussed for 
fully relativistic particles can be
 applied to the case of multiple production in the two bodies 
 collision at high energy. As it is well 
known, the secondary particles produced in these experiments have a 
transverse momentum
 $p_t$, which is negligible compared with the momentum parallel to 
the  collision axis
 ($p_l$), and in fact, one-dimensional variables have been introduced
 already to describe the process\cite{fey}. 
A very popular phenomenological model considers the secondaries in the 
final state as non-interacting particles grouped in clusters
having a classical statistical 
distribution in the longitudinal momentum or rapidity\cite{bal}.
 Let us think of a cluster in terms of our model (adapted for the purpose)
 and  consider the relative momenta of the secondaries emitted from it
 to have values $ \sim p_t$,
 which is much less (outside the central zone) than the total momentum of 
the system. Then we
 must conclude that the reduced density matrix of clusters 
 rapidly diagonalizes on the 
basis of their total momentum. Which justifies their description in terms of 
classical statistical distribution.

       \vskip 1.4 true cm

\end{document}